\RequirePackage{ifpdf}
\ifpdf 
\documentclass[pdftex]{sigma}
\else
\documentclass{sigma}
\fi

\def\R{\mathbb{R}}

\def\R{\mathbb{R}}
\def\Z{\mathbb{Z}}

\def\r{\rangle}
\def\l{\langle}

\begin{document}

\renewcommand{\PaperNumber}{025}

\FirstPageHeading

\ShortArticleName{Compact Simple Lie Groups  and Their $C$-, $S$-, and
$E$-Transforms}

\ArticleName{Compact Simple Lie Groups\\
 and Their $\boldsymbol{C}$-, $\boldsymbol{S}$-, and
$\boldsymbol{E}$-Transforms}

\Author{Jiri PATERA} 
\AuthorNameForHeading{J. Patera}

\Address{Centre de Recherches Math\'ematiques,
         Universit\'e de Montr\'eal,\\
         C.P.6128-Centre ville,
         Montr\'eal, H3C\,3J7, Qu\'ebec, Canada}
\Email{\href{mailto:patera@crm.umontreal.ca}{patera@crm.umontreal.ca}}
\URLaddress{\url{http://www.crm.umontreal.ca/~patera/}}

\ArticleDates{Received December 01, 2005; Published online December 09, 2005}

\Abstract{New continuous group transforms, together with their
discretization over a~lattice of any density and admissible
symmetry, are defined for a general compact simple Lie groups of
rank $2\leq n<\infty$. Rank 1 transforms are known. Rank 2
exposition of $C$- and $S$-transforms is in the literature. The
$E$-transforms appear here for the first time.}


\Classification{22E46; 33C80; 43A80; 94A08}

\section{Introduction}

In the talk \cite{Ptalk} three types of transforms, called  
$C$-, $S$-, and $E$-transforms, are introduced for each compact
simple Lie group $G$. Number of variables in the transforms equals
the rank of $G$. Generalization from simple to semisimple Lie
groups is straightforward but we disregard it here.

The transforms are defined on a finite region $F$ of a real
$n$-dimensional Euclidean space $\R^n$, more precisely $F$ is a
simplex often called the fundamental region of $G$. The
transforms are introduced simultaneously as continuous ones on
$F$, and also as discrete transforms on a lattice grid
$F_M\subset F$ of any density fixed by a positive integer $M$.
Respectively the three transforms are multidimensional
generalizations of the cosine transform, the sine transform and
the common Fourier transform using exponential functions.

A practical motivation for studying the group transforms introduced
here, comes from the abundance of multidimensional digital data
which need to be processed according to diverse criteria. In
particular, continuous extension of discrete transforms, using
$C$-functions, appears to be rather simple and advantageous way to
interpolate digital data. It was noticed as a~property of
$C$-transforms  based on $G=SU(2)$ and $SU(2)\times SU(2)$, in
\cite{AP1}, although a crucial step toward that was made already in
\cite{Wa,Ag}. There is every reason to expect that the same is true
about the $S$- and $E$-transforms.  Moreover, the general approach
described here allows one to treat not only the data in 2 or 3
dimensions, but in any dimension $n<\infty$ using any of the
semisimple Lie groups of rank $n$.

General discrete $C$-transforms are the content of \cite{MP3}.
Their applications are in \cite{MP2,MMP1,MMP2,GP}. In \cite{PZ1,PZ2}
one finds an explicit description of the continuous and discrete
$C$-transforms for the four semisimple Lie groups of rank 2. For
other applications see  \cite{AP1,APSA} and \cite{P1,P2,AP3}.
Let us also point out a forthcoming review of a $C$-functions~\cite{KP}.

The $S$-transform for rank 2 groups are the content of~\cite{PZ3}.

The $E$-transforms are not found in the literature sofar~\cite{KPZ}.

\section{General discrete and continuous transforms in $n$
dimensions}

There is an underlying compact semisimple Lie group $G$ implied in
the considerations here. Let $\Phi$ denote the new expansion
functions of either type, $C$ or $S$, or $E$. Within each family
the functions are orthogonal when integrated over $F$. The
following relations are called the continuous transform and its
inversion:
\begin{gather}\label{transform}
f(x)=\sum_\lambda f_\lambda\Phi_\lambda(x),
\qquad
f_\lambda=\int_F f(x)\overline{\Phi_\lambda(x)}\,dx
\end{gather}
assuming a suitable normalization of the expansion functions.
Here $x\in\R^n$ stands for $n$ continuous variables, $\lambda\in
P^+$ is a point of an $n$ dimensional lattice $P$, which is taken
to be in the `positive chamber' $P^+$ of $P$ whenever it is
convenient. The finite region $F\subset\R^n$ is the fundamental
region of the appropriate Weyl group for $C$- and
$S$-transform, and it is a pair of adjacent copies of $F$ in case
of the $E$-transform. Overbar denotes complex conjugation.

Note that $\Phi$ in \eqref{transform} could be the irreducible
character of $G$. Practical advantage of any of the three
families of functions here is in that they are much simpler than
the irreducible characters.

Discretization of the transform \eqref{transform} and its
inversion, namely 
\begin{gather}\label{dtransform}
f(x_k)=\sum_{\lambda\in S_M} f_\lambda\Phi_\lambda(x_k),
\qquad
f_\lambda=\sum _{x_k\in F_M}
        f(x_k)\overline{\Phi_\lambda(x_k)},
\end{gather}
require two modifications of the continuous case. Firstly the
continuous variables $x$ need to be replaced by a suitable grid
$F_M$ of discrete points in $F$. Secondly, over the finite set
$F_M$ of points only a finite number of functions $\Phi_\lambda$,
given by $\lambda\in S_M$, can be pairwise orthogonal and thus
participate in the expansion \eqref{dtransform}. Density of the
points in $F_M$ is fixed by a positive integer $M$.

In order to define the transform \eqref{transform}, one needs to
choose the group $G$ (for simplicity of formulation here, $G$ is
supposed to be simple), and to define the appropriate set of
orthogonal and distinct expansion functions.

In order to define the discrete transform \eqref{dtransform}, one
needs as well to choose the sets $F_M$ and~$S_M$. For a fixed
value of~$M$, the set $F_M$ is unique, while $S_M$ is not. However
there is always a unique lowest set $S_M$; we assume that one to
be always chosen.

\section{Preliminaries}

Standard Lie theory provides description of the following objects
for any compact simple Lie group $G$ of rank $1\leq n<\infty$:
\begin{enumerate}
\itemsep=0pt
\item[]
maximal torus $T$; the root system and its subset of simple roots
$\{\alpha_1,\dots,\alpha_n\}$; finite Weyl group $W$ of $G$; the
weight lattice $P$ and its dual $\hat P$ in the real Euclidean space
$\R^n$ of dimension $n$; the bases $\{\omega_1,\dots,\omega_n\}$ and
$\{\hat\omega_1,\dots,\hat\omega_n\}$ of $P$ and $\hat P$
respectively; the Weyl group orbit $W_\lambda$ of $\lambda\in P$; the
size $|W_\lambda|$ of $W_\lambda$; the fundamental region $F$ for $W$
action on~$T$; finite $W$-invariant subgroup $A_M\subset T$
generated by the elements of order $M$ in $T$.
\end{enumerate}

A description of the points $x_s\in F_M=A_M\cap F$ is given next
\begin{gather}\label{grid}
F_M= \left\{x_s=\sum_{k=1}^n\frac{s_k}M\hat\omega_k\,\Big|\,
     s_k\in\Z^{\geq0};\ M=s_0+\sum_{m=1}^nq_ms_m\right\},
\end{gather}
where the coefficients $q_m$ are given in the highest root $\xi$
of $G$:
\[
\xi=\sum_{m=1}^nq_m\alpha_m.
\]
To find the points of $F_M$, it suffices to find all the
non-negative integers $\{s_0,s_1,\dots,s_n\}$ that add up to $M$
according to \eqref{grid}. It is an easy computing task. In the case
of $E$-functions one has to take into account that $F^e$ consists of
two copies of $F$.

More theoretically, points of $F_M$ are representatives of
conjugacy classes of elements of an~Abelian subgroup $A\subset T$
of the maximal torus $T$, which is generated by all elements of
order~$M$ in $T$. Thus one has $F_M=A\cap F$.

\section[Definition of C-functions]{Definition of $\boldsymbol{C}$-functions}

A review of properties of $C$-functions denoted here
$C_\lambda(t)$ is in the forthcoming paper \cite{KP}
\begin{gather}\label{C}
C_\lambda(t):=\sum_{\lambda'\in W_\lambda}
  e^{2\pi i\l\lambda'\mid t\r},\qquad
\lambda\in P^+\subset\R^n,\quad t\in\R^n.
\end{gather}
Here $\l\lambda'\mid t\r$ denotes the scalar product in $\R^n$.
The number of summands is finite, it is equal to the size of the
Weyl group orbit of $\lambda$.

Among the useful properties of $C$-functions, note the complete
decomposition of products
\[
C_\lambda(t)C_{\lambda'}(t)=C_{\lambda+\lambda'}(t)+C_
\mu(t)+\cdots.
\]
The functions are continuous and have all derivatives continuous in
$\R^n$. They are $W$-invariant and have interesting symmetry
properties \cite{PZ1,PZ2} with respect to affine $W$. In
particular, they are symmetric with respect to reflection in the
sides of $F$ of maximal dimension, i.e.\ $n-1$. Hence their normal
derivative at the boundary is zero.

  One has the continuous orthogonality of $C$-functions,
\begin{gather}\label{Corthog}
\left(C_\lambda,C_\mu\right)=\int_F
C_\lambda(t)\overline{C_\mu(t)}dt
               \sim\delta_{\lambda\mu},
\end{gather}
and the discrete orthogonality of $C$-functions,
\[
\sum_{t\in A\cap F} |Wt|C_\lambda(t)\overline{C_\mu(t)}\sim
\delta_{\lambda\mu},
\]
where the integer $|Wt|$ is the size of the $W$-orbit of $t$. The
lattice points $\lambda,\mu\in P^+$ are subjects to additional
restriction assuring that $C_\lambda(t)$, $C_\mu(t)$ belong to a
finite set of the functions denoted~$S_M$. The Abelian
$W$-invariant group $A$ can be built as the group generated by
representatives of the conjugacy classes of elements of given
order $M<\infty$ in the Lie group. For the rank~2 cases, see~\cite{PZ1,PZ2}.

\section[Definition of S-functions]{Definition of $\boldsymbol{S}$-functions}

Comparing with \eqref{C}, we have
\begin{gather}\label{S}
S_\lambda(t):=\sum_{\lambda'\in W_\lambda}
  (-1)^{l(\lambda,\lambda')} e^{2\pi i\l\lambda'\mid t\r},\qquad
\lambda\in P^{++}\subset\R^n,\quad t\in\R^n.
\end{gather}
Here $l(\lambda,\lambda')$ is the minimal number of elementary
reflections from $W$ needed to transform $\lambda$ into
$\lambda'$; $P^{++}$ denotes the interior of the positive chamber
$P^+$ of the weight lattice $P$. The $S$-functions are continuous
and antisymmetric with respect to reflection in the sides of $F$ of
maximal dimension, i.e.\ $n-1$. Hence their value at the boundary
is zero.

A product of an even number of $S$-functions decomposes into the
sum of $C$-functions where coefficients are positive and
negative integers. A product of $C$- and $S$-functions decomposes
into the sum of $S$-functions:
\begin{gather}
S_\lambda(t)S_{\lambda'}(t)=C_{\lambda+\lambda'}(t)-C_\mu(t)
+\cdots,\\
C_\lambda(t)S_{\lambda'}(t)=S_{\lambda+\lambda'}(t)+S_\mu(t)+\cdots.
\end{gather}
See some examples in \cite{PZ3}.

Continuous orthogonality of $S$-functions formally coincides
with that for $C$-functions. The only difference is in the fact
that $\lambda\in P^{++}$ rather than in $P^+$.
\begin{gather}
\left(S_\lambda,S_\mu\right)=\int_F
S_\lambda(t)\overline{S_\mu(t)}dt
               \sim\delta_{\lambda\mu}.
\end{gather}

Discrete orthogonality of $S$-functions happens to be on the same
grids in $F$ as in the case of $C$-functions. Since the
$W$-orbit of all the points in the interior $P^{++}$ of the
positive chamber $P^+$ is the same,
\[
\sum_{t\in A\cap F}S_\lambda(t)\overline{S_\mu(t)}
\sim\delta_{\lambda\mu}.
\]
As before, one needs to assume that $\lambda$ and $\mu$ belong to
the lowest finite set, denoted $S_M$, of dominant weights of
pairwise orthogonal functions over the grid fixed by the positive
integer $M$. For that some additional restrictions on  $\lambda$ and
$\mu$ need to be imposed.

\section[Definition of E-functions]{Definition of $\boldsymbol{E}$-functions}

In order to define the $E$-functions in a way analogous to the
$C$- and $S$-functions, one needs to replace the Weyl group $W$
by its even subgroup $W^e\subset W$ and correspondingly enlarge
its fundamental region $F^e$.

Let $r$ be any simple reflection from the Weyl group. Then
\[
F^e:= F\cup rF,
\]
where $F^e$ is the fundamental region for $W^e$.

For $\lambda\in P$,
\begin{gather}
E_\lambda(t):=\sum_{\lambda'\in W^e_\lambda}
  e^{2\pi i\l\lambda',t\r}.
\end{gather}

Since $E_\lambda(t)$ depends on $W^e\lambda$, not on $\lambda$, we
can suppose $\lambda\in P^{+e}=P^+\cup rP^+$. One verifies
directly that
\begin{gather}
C_\lambda=\left\{
 \begin{array}{ll}
    E_\lambda+E_{r\lambda}\ \ &\text{if}\ \ \lambda\neq r\lambda,\\
    E_\lambda                  &\text{if}\ \ \lambda=r\lambda,
 \end{array}\right.
\qquad
S_\lambda=\left\{
 \begin{array}{ll}
    E_\lambda-E_{r\lambda}\ \ &\text{if}\ \ \lambda\neq r\lambda,\\
    0                  &\text{if}\ \ \lambda=r\lambda.
 \end{array}\right.
\end{gather}

\medskip

Continuous orthogonality of $E$-functions involves integration
over $F^e$: 
\begin{gather}
\left(E_\lambda,E_\mu\right)
  =\int_{F^e} E_\lambda(t)\overline{E_\mu(t)}dt
  \sim\delta_{\lambda\mu}.
\end{gather}

Discrete orthogonality of $E$-functions can be introduced in a
similar way as for the $C$- and $S$-functions, using the same grid
of points on $F^e$. 

Let $A\subset\hat{T}$ be a finite subgroup which is $W^e$-invariant.
Again a convenient and versatile set up would be to take the group
generated by representatives of conjugacy classes of elements of a
fixed order $M$ in the Lie group. The intersection of $A$ with $F^e$
is a finite set of points
\[
A\cap F^e=\{z_1^e,z_2^e,\dots,z_m^e\},\qquad
 o_j^e=|W^ez_j^e|.
\]

Since $E_\lambda$, $E_\mu$ are $W^e$-invariant, we have
\[
\sum_{j=1}^m o_j^eE_\lambda(z_j^e)\overline{E_\mu(z_j^e)}
   \sim\delta_{\lambda\mu}.
\]

As in the previous cases, only a finite set of $E$-functions can be
pairwise orthogonal and distinct on a finite grid of points given by
fixed value of $M$. Therefore some additional restrictions on
$\lambda$, $\mu$ need to be imposed.

\begin{figure}
\centering
\includegraphics[width=12cm]{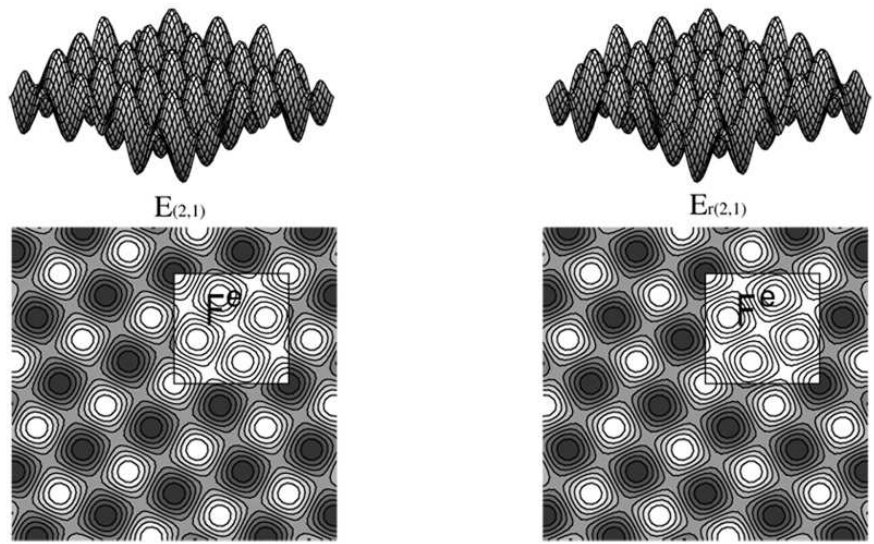}
\caption{3-D and contour graphs of $E_{(2,1)}$ and $E_{r(2,1)}$ for
Lie group $C_2$. The light color square is the fundamental region
$F^e$ of $C_2$.} \label{fig}
\end{figure}

Some two dimensional examples can be found in the forthcoming
\cite{KPZ}. Fig.~\ref{fig} shows an example of $E_{(2,1)}$ and
$E_{r(2,1)}$ for Lie group $C_2$.

\subsection*{Acknowledgements}

The author is grateful for partial support for the work from the
National Science and Engineering Research Council of Canada and from
MITACS. The author is also grateful to A.~Zaratsyan for providing
the figures of $E$-functions.

\LastPageEnding


\begin{thebibliography}{99}
\footnotesize

\bibitem{Ptalk}
Patera J., Invited talk at the Sixth International Conference
``Symmetry in Nonlinear
Mathematical Physics'' (June 20--26, 2005, Kyiv).

\bibitem{AP1}
Atoyan A., Patera J., Properties of continuous Fourier extension
of the discrete cosine transform and its multidimensional
generalization, {\it J. Math. Phys.}, 2004, V.45, 2468--2491.

\bibitem{Wa}
Wang Zhongde, Interpolation using type I discrete cosine
transform, {\it Electron. Lett.}, 1990, V.26, 1170--1171.

\bibitem{Ag}
 Agbinya J.I., Two dimensional interpolation of real
sequences using the DCT, {\it Electron. Lett.}, 1993, V.29, 204--205.

\bibitem{MP3}
Moody R.V., Patera J., Computation of character
decompositions of class  functions on compact semisimple Lie
groups, {\it Math. Comp.}, 1987, V.48, 799--827.


\bibitem{MP2}
Moody R.V., Patera J.,
Elements of finite order in Lie groups and their applications,
XIII Int. Colloq. on Group Theoretical Methods in Physics,
Editor W.~Zachary,  Singapore, World Scientific Publishers, 1984,
308--318.

\bibitem{MMP1}
McKay W.G., Moody R.V., Patera J.,
Decomposition of  tensor products of $E_8$ representations,
{\it 
Algebras Groups Geom.}, 1986, V.3, 286--328.

\bibitem{MMP2}
McKay W.G., Moody R.V., Patera J., Tables of $E_8$
characters and decomposition of plethysms, in Lie Algebras
and  Related Topics, Editors D.J.~Britten,  F.W.~Lemire and R.V.~Moody,
Providence R.I., Amer. Math. Society,  1985, 227--264.

\bibitem{GP}
 Grimm S., Patera J.,  Decomposition of tensor products of
the fundamental  representations of $E_8$, in Advances in
Mathematical Sciences:  CRM's 25 Years, Editor L.~Vinet, {\it CRM
Proc. Lecture Notes},  Providence RI, Amer. Math. Soc.,  1997, V.11, 
329--355.

\bibitem{PZ1}
 Patera J., Zaratsyan A.,  Cosine transform generalized to Lie
groups $SU(2)\times SU(2)$ and $O(5)$, {\it J. Math. Phys.}, 2005, V.46,
053514, 25 pages.

\bibitem{PZ2}
Patera J., Zaratsyan A., Cosine transform generalized to Lie
groups $SU(3)$ and $G(2)$, {\it J. Math. Phys.}, 2005, V.46, 113506, 17 pages.

\bibitem{APSA}
Atoyan A., Patera J., Sahakian V., Akhperjanian A., 
Fourier transform method for imaging atmospheric Cherenkov
telescopes, {\it Astroparticle Phys.}, 2005, V.23,
79--95.

\bibitem{P1}
 Patera J.,  Orbit functions of compact semisimple Lie groups
as special functions, in Proceedinds of Fifth International Conference ``Symmetry in Nonlinear
Mathematical Physics'' (June 23--29, 2003, Kyiv),
Editors A.G.~Nikitin, V.M.~Boyko, R.O.~Popovych and I.A.~Yehorchenko, {\it Proceedings of Institute
of Mathematics}, Kyiv, 2004, V.50, Part~3, 1152--1160.

\bibitem{P2}
 Patera J., Algebraic solutions of the Neumann boundary value
problems  on fundamental region of a compact semisimple Lie
group, talk given at the Workshop on Group Theory and
Numerical Methods (May 26--31, 2003, Montreal).

\bibitem{AP3}
 Atoyan A., Patera J., Continuous extension of the discrete
cosine transform, and its   applications to data processing,
Proceedings of the Workshop on Group Theory  and Numerical Methods 
(May 26--31, 2003, Montreal).

\bibitem{KP}
 Klimyk A., Patera J., Orbit functions, in preparation.


\bibitem{PZ3}
 Patera J., Zaratsyan A., The sine transform generalized to
semisimple Lie groups rank 2, Preprint, 2005.

\bibitem{KPZ}
 Kashuba I., Patera J., Zaratsyan A., The $E$-functions of
compact semisimple Lie groups and their discretization, in
preparation.

\bibitem{BMP}
Bremner M.R., Moody R.V., Patera J., Tables of dominant weight
multiplicities for representations of simple Lie algebras, New
York, Marcel Dekker,  1985, 340~pages.


\bibitem{MP4}
Moody R.V., Patera J., Discrete and continuous orthogonality
of $C$-, $S$-, and $E$-functions of a compact semisimple Lie
group, in preparation.

\end{thebibliography}
\end{document}